\documentclass[sigconf]{acmart}

\AtBeginDocument{%
  \providecommand\BibTeX{{%
    \normalfont B\kern-0.5em{\scshape i\kern-0.25em b}\kern-0.8em\TeX}}}

\copyrightyear{2024}
\acmYear{2024}
\setcopyright{acmlicensed}\acmConference[ICSE '24]{2024 IEEE/ACM 46th
International Conference on Software Engineering}{April 14--20, 2024}{Lisbon,
Portugal}
\acmBooktitle{2024 IEEE/ACM 46th International Conference on Software
Engineering (ICSE '24), April 14--20, 2024, Lisbon, Portugal}
\acmDOI{10.1145/3597503.3639194}
\acmISBN{979-8-4007-0217-4/24/04}

\acmSubmissionID{icse24-main-2202}

\acmConference[ICSE 2024]{46th International Conference on Software Engineering}{April 2024}{Lisbon, Portugal}

\newcommand{\nd}{\vspace{1mm}\noindent}

\usepackage[shortlabels]{enumitem}

\usepackage{pifont}
\usepackage{forest}

\newcommand{\circled}[1]{\tikz[baseline=(char.base)]{
    \node[shape=circle, draw=black, fill=black, text=white, inner sep=1pt] (char) {\small\strut #1};}}

\newcommand{\percentagebar}[1]{\rule{#1}{6pt}}
\usepackage{fontawesome5}

\newcommand{\cmark}{\ding{51}}%
\newcommand{\xmark}{\ding{55}}%

\usepackage{marvosym}

\newcommand*{\usericon}{\includegraphics[scale=0.17]{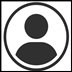}}
\newcommand*{\chatgpticon}{\includegraphics[scale=0.17]{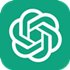}}
\usepackage[many]{tcolorbox}%

\let\oldquote\quote
\let\endoldquote\endquote
\renewenvironment{quote}[2][]
  {\if\relax\detokenize{#1}\relax
     \def\quoteauthor{#2}%
   \else
     \def\quoteauthor{#2~---~#1}%
   \fi
   \oldquote}
  {\par\nobreak\smallskip\hfill(\quoteauthor)%
   \endoldquote\addvspace{\bigskipamount}}
\usepackage[noorphans,vskip=0.5ex]{quoting}
\usepackage{soul}

\newcommand{\jRevision}[1]{\textcolor{black}{{#1}}}
\newcommand{\rev}[1]{\textcolor{black}{#1}}

\begin{document}

\def\bf{\textbf}
\def\ib{\textit{\textbf}}
\def\eq {Equation~}
\def\eqm {Eq~}
\def\eqs {Equations~}
\def\fig {Figure~}
\def\figs {Figures~}
\def\tbl {Table~}
\def\tbls {Tables~}
\def\ie{\textit{i.e.,}}
\def\eg{\textit{e.g.,}}
\def\sec {Section~}
\def\secs {Sections~}
\def\alg {Algorithm~}
\def\algs {Algorithms~}
\def\app {Appendix~}
\def\it{\textit}
\def\tr{\textrm}
\def\tt{\mct}
\title{ChatGPT Incorrectness Detection in Software Reviews}

\author{Minaoar Hossain Tanzil}
\email{minaoar@gmail.com}
\orcid{0000-0002-3323-4917}
\affiliation{%
  \institution{University of Calgary}
  \city{Calgary}
  \state{Alberta}
  \country{Canada}
}

\author{Junaed Younus Khan}
\email{junaedyounus.khan@ucalgary.ca}
\orcid{0000-0001-8138-1105}
\affiliation{%
  \institution{University of Calgary}
  \city{Calgary}
  \state{Alberta}
  \country{Canada}
}

\author{Gias Uddin}
\email{guddin@yorku.ca}
\orcid{0000-0003-1376-095X}
\affiliation{%
  \institution{York University}
  \city{Toronto}
  \state{Ontario}
  \country{Canada}
}

\renewcommand{\shortauthors}{Tanzil et al.}

\begin{abstract}
We conducted a survey of 135 software engineering (SE) practitioners to understand how they use Generative AI-based chatbots like ChatGPT for SE tasks. We find that they want to use ChatGPT for SE tasks like software library selection but often worry about the truthfulness of ChatGPT responses. We developed a suite of techniques and a tool called CID (ChatGPT Incorrectness Detector) to automatically test and detect the incorrectness in ChatGPT responses. CID is based on the iterative prompting to ChatGPT by asking it contextually similar but textually divergent questions (using an approach that utilizes metamorphic relationships in texts). The underlying principle in CID is that for a given question, a response that is different from other responses (across multiple incarnations of the question) is likely an incorrect response. In a benchmark study of library selection, we show that CID can detect incorrect responses from ChatGPT with an F1-score of 0.74 - 0.75.
\end{abstract}



\maketitle
\balance

\section{Introduction}
Thanks to the recent advent of pre-trained large language models (LLMs) like ChatGPT \cite{ChatGPT_openAI}, many domains are now undergoing rapid automation transformation. The interactive interface of ChatGPT entices users of different professions to use it as a personal assistant and software engineering (SE) is no exception. Studies employed LLMs in various SE tasks like code generation \cite{finnie2022robots, khan2023combining, liu2023your}, program repair \cite{xia2023keep, prenner2021automatic, pearce2023examining}, and code summarization/explanation \cite{khan2022automatic, ahmed2022few, macneil2022generating}. However, LLMs can generate incorrect results \cite{feldman2023trapping, zhang2023language, bang2023multitask, galitsky2023truth}. 

In this paper, we conduct a survey and develop a tool to assess the quality of ChatGPT while supporting SE practitioners. We focus on software library selection as a case study for which developers would like to use ChatGPT, because we know that developers rely on online reviews during library selection~\cite{uddin2019understanding}. 

\ul{In the first phase, we surveyed 135 SE practitioners to learn their preferences and concerns about ChatGPT usage}. We find that
\begin{itemize}[leftmargin=9pt,nolistsep]
\setlength\itemsep{0em}
\item SE practitioners are enthusiastic to use ChatGPT for diverse SE applications, such as code generation, library exploration, etc. 
\item However, most of them do not fully rely on the responses and consider them as unreliable or somewhat reliable. 
\item Developers employ various manual but cumbersome strategies to verify ChatGPT responses like further searches in Google or Stack Overflow, seeking additional clarifications from ChatGPT through follow-up questions, etc. They wished for automated tool support to assess the inaccuracies in ChatGPT responses.
\end{itemize}

\ul{In the second phase, we develop a suite of techniques and a tool called CID (ChatGPT Incorrectness Detector) that can automatically determine inaccuracies in ChatGPT responses}. 

One might argue that the logistic probability of the tokens in LLMs can be indicative of the correctness of the generated results \jRevision{\cite{varshney2023stitch}}. However, in LLMs, log probabilities mainly indicate uncertainty over specific tokens, rather than capturing epistemic uncertainty related to the correctness of the given information \jRevision{\cite{lin2022teaching}}. When a claim can be paraphrased in various ways, each paraphrase may have a low log probability regardless of the epistemic correctness. Lin et al. showed that such models can be calibrated to express their level of epistemic uncertainty (namely, verbalized probability) which, however, requires domain-specific fine-tuning \cite{lin2022teaching}. 
It is also reported that though the base pre-trained model of GPT-4 (which happens to be the underlying model of ChatGPT) is highly calibrated, the calibration is affected by further post-training process i.e., Reinforcement Learning from Human Feedback (RLHF) \cite{openaiGPT4}. Moreover, all of these approaches are gray/white-box testing, because they either require the original model or its underlying architectural details. Such approaches are not suitable for scenarios when people use third-party chatbots like ChatGPT. 

 In contrast, for our CID tool, we derive a black-box testing approach. Our approach is based on iterative but automated guided prompting in ChatGPT, where the prompted questions can be mutated without changing the base query, i.e., what we are asking across multiple questions for a given prompt remains the same, but we ask the questions in different ways. We employ metamorphic relationships to mutate the questions. We then calculate the similarity of the generated responses and use the similarity assessment to automatically label whether a response is incorrect. The underlying principle in CID is that for an original ChatGPT response that is incorrect, the subsequent question responses are likely to be (more) inconsistent (i.e., dissimilar). In a benchmark study of software library selection, we show that CID can detect such incorrect responses with an F1-score of 0.74 - 0.75.
\section{Survey to Assess Software Developer Perspectives on ChatGPT Usage}
We surveyed 135 SE practitioners to understand how they use generative artificial intelligence (AI) based chatbots like ChatGPT for SE tasks. We answer three research questions (RQs):
\begin{enumerate}[label=\textbf{RQ\arabic{*}.}, leftmargin=30pt]
    \item  Why do software developers use ChatGPT?
    \item How much do developers rely on ChatGPT responses?
    \item How do developers verify ChatGPT responses?
\end{enumerate}
\subsection{Survey Setup}
\rev{The surveys were conducted after following review and approval from the Research Ethics Board of the associated university.}
\subsubsection{Survey Questions}
We asked each participant 10 questions listed in Table \ref{tab:survey_questions_on_chatgpt}. We designed the survey in a semi-close-ended fashion where most questions had the opportunity to write down additional feedback. \rev{The survey questionnaire was designed based on a pilot survey. In the pilot survey, we provided the questionnaire to 24 industry professionals from big software companies across the world. The recruitment of these professionals was conducted by mostly convenient sampling from the professional network of the authors. Before distributing the survey, we provided them with a short briefing of our intention. Finally, 20 professionals participated in the survey. Based on their survey responses and post-survey feedback, we re-adjusted our final survey}.

\subsubsection{Survey Participants}
\begin{table}[t]
  \centering
  \caption{Demography of Survey Participants}
  \label{tab:demography_of_survey_participants}
  \resizebox{\columnwidth}{!}{%
    \begin{tabular}{rcccccc}
      \textbf{} & \multicolumn{6}{c}{\textbf{Years of Experience}} \\
      \cline{2-7}
      \multicolumn{1}{r|}{\textbf{Current Profession}} & \textbf{0-5} & \textbf{6-10} & \textbf{11-15} & \textbf{16-20} & \multicolumn{1}{c|}{\textbf{21-24}} & \textbf{Total} \\
      \hline
      \multicolumn{1}{r|}{\begin{tabular}[c]{@{}r@{}}\textbf{Software   Engineer}\\ (Dev/QA/Data/Ops)\end{tabular}} & 64 & 34 & 11 & 4 & \multicolumn{1}{c|}{-} & \textbf{113} \\
      \hline
      \multicolumn{1}{r|}{\begin{tabular}[c]{@{}r@{}}\textbf{Manager}\\ (Dev/PM/Ops)\end{tabular}} & {-} & 3 & 10 & 2 & \multicolumn{1}{c|}{-} & \textbf{15} \\
      \hline
      \multicolumn{1}{r|}{\begin{tabular}[c]{@{}r@{}}\textbf{Executive}\\ (Company Leadership)\end{tabular}} & {-} & {-} & 1 & 1 & \multicolumn{1}{c|}{2} & \textbf{4} \\
      \hline
      \multicolumn{1}{r|}{\begin{tabular}[c]{@{}r@{}}\textbf{Non-Tech}\\ (Sales/Marketing/CS/HR)\end{tabular}} & {-} & 2 & 1 & {-} & \multicolumn{1}{c|}{-} & \textbf{3} \\
      \hline
      \multicolumn{1}{r|}{\textbf{Total}} & \textbf{64} & \textbf{39} & \textbf{23} & \textbf{7} & \multicolumn{1}{c|}{\textbf{2}} & \textbf{135} \\
    \end{tabular}%
  }
\end{table}
The participants' roles include software engineers, developers, managers, etc. Table \ref{tab:demography_of_survey_participants} shows the distribution of the participants over their profession and years of experience. \rev{We used the snowball sampling~\cite{Goodman-SnowballSampling-AnnMathStat1961} approach to select the participants}. First, we picked 12 senior individuals (from our personal contacts) having at least 15 years of experience in the industry. We recorded their feedback and also requested them to invite appropriate candidates. The second batch of participants included 123 individuals ranging from developers to managers. We deliberately avoided open social or professional networking sites as a means of participant recruitment. While such platforms may offer access to a large number of potential participants, the lack of control over the respondents' qualifications and expertise could compromise the reliability of our study. \rev{The 20 participants from the pilot survey were omitted from our list of 135 survey participants}.

\begin{table*}
\caption{Survey questions and  their mapping to
the Research Questions. Here, C/O=Close/Open-ended question, G/S=Generic/Scenerio-based question. For scenario-based questions, we used library selection as a case-study.}
\label{tab:survey_questions_on_chatgpt}
\begin{tabular}{c l c c c}
\hline
\textbf{Q\#} & \textbf{Questions}                                                                                          & \textbf{O/C} & \textbf{G/S} & \textbf{RQ} \\ \hline
1            & Did you use ChatGPT?                                                                                        & C            & G            & 1.1         \\
2            & In general, which of the cases you used it for?                                                           & C            & G            & 1.1         \\
3            & As a software professional, how did you or can you use it?                                                  & C            & G            & 1.1         \\
4            & How would you describe your experience with using it so far?                                                & C            & G            & 1.1         \\
5            & How much do you rely on the content/response of ChatGPT?                                            & C            & G            & 1.2         \\
6            & Have you considered using ChatGPT to select or compare software libraries? Please share the pros and cons. & O            & S            & 1.2         \\
7            & How much would you rely on ChatGPT's response for the given library selection query?                        & C            & S            & 1.3         \\
8            & Would you rely on the ChatGPT's response after further inquiry?                                             & C            & S            & 1.3         \\
9            & Do you think the opinion from ChatGPT is correct?                                                           & C            & S            & 1.3         \\
10           & What can be the ways to improve the reliability of ChatGPT responses?                                       & C            & G            & 1.3       \\ \hline
\end{tabular}%
\vspace{-4mm}
\end{table*}

\subsection{Reasons for using ChatGPT (RQ1)}
\nd\circled{Q1} We asked the participants whether they have used ChatGPT or not. The vast majority (133 respondents) reported having already used ChatGPT, while only 2 participants responded in the negative.
\begin{itemize}[leftmargin=9pt,nolistsep]
    \setlength\itemsep{0em}
    \item {\textit{\textbf{Yes:}}~\color{black!90}\percentagebar{49pt}} \textbf{98.52\%} 
    \item {\textit{\textbf{No:}}~\color{black!90}\percentagebar{2pt}} \textbf{1.48\%} 
\end{itemize}

\nd\circled{Q2} We were interested in understanding the common purposes ChatGPT serve to the software professionals regardless of their specific roles in software engineering. After all, they are human too. As this was a multiple-choice checklist question, respondents were allowed to select multiple options.

\begin{itemize}[leftmargin=9pt,nolistsep]
    \setlength\itemsep{0em}
    \item {\textit{\textbf{Just for Fun:}}~\color{black!90}\percentagebar{21pt}} \textbf{42.22\%\\}
    57 participants used ChatGPT just for fun and entertainment purposes, engaging with its capabilities in a playful manner.
    \item {\textit{\textbf{As a Search Engine:}}~\color{black!90}\percentagebar{39pt}} \textbf{79.26\%\\}
    The majority (107) of respondents utilized ChatGPT as a search engine to collect information, highlighting its potential as an AI-powered information retrieval tool.
    \item {\textit{\textbf{Content Creation:}}~\color{black!90}\percentagebar{23pt}} \textbf{46.67\%\\}
    63 participants used ChatGPT to generate new content, such as presentations, documents, images, and more.
    \item {\textit{\textbf{Decision Making:}}~\color{black!90}\percentagebar{24pt}} \textbf{48.15\%\\}
    65 respondents used ChatGPT for decision-making purposes.
    \item {\textit{\textbf{Exploration and Data Collection}:}~\color{black!90}\percentagebar{13pt}} \textbf{27.41\%\\}
    37 participants employed ChatGPT for initial exploratory data collection, utilizing its capabilities to gain insights into a topic.
    \item {\textit{\textbf{Research for Product/Business Development:}}~\color{black!90}\percentagebar{14pt}} \textbf{28.15\%\\}
    38 participants used ChatGPT for market research, specifically to explore new products or business development ideas.
    \item {\textit{\textbf{Learning and Knowledge Acquisition: }}~\color{black!90}\percentagebar{31pt}} \textbf{61.48\%\\}
    A significant proportion (83) of participants used ChatGPT as an educational resource to learn about new technologies.
    \item {\textit{\textbf{Others: }}~\color{black!90}\percentagebar{2pt}} \textbf{1.32\%\\}
    A few respondents mentioned some additional use cases for ChatGPT such as asking questions, generating ideas, and finding solutions to different problems. However, these use cases can also be fitted into the above-predefined options. 

\end{itemize}

\nd\circled{Q3} We asked the developers how they (are willing to) use ChatGPT in their work as software professionals. The question was a multiple-choice checklist. 

\begin{itemize}[leftmargin=9pt,nolistsep]
    \setlength\itemsep{0em}
    \item {\textit{\textbf{Code Generation and Optimization:}}~\color{black!90}\percentagebar{30pt}} \textbf{60\%\\}
    81 survey participants use ChatGPT for generating code for their applications, demonstrating its value in code completion, generation, optimization, and refactoring tasks.
    \item {\textit{\textbf{Code Analysis and Review:}}~\color{black!90}\percentagebar{26pt}} \textbf{52.59\%\\}
    71 respondents leverage ChatGPT for analyzing code-related tasks, such as code comparison, debugging, summarization, and code review. This highlights ChatGPT's significance in assisting developers with code understanding and error identification.
    \item {\textit{\textbf{Problem-Solving Support:}}~\color{black!90}\percentagebar{44pt}} \textbf{86.67\%\\}
    A significant proportion of our survey participants (117) turn to ChatGPT to find solutions to coding problems they encounter. This indicates that ChatGPT is a valuable resource for developers when facing challenges during the development process.
    \item {\textit{\textbf{Exploring Alternative Approaches:}}~\color{black!90}\percentagebar{37pt}} \textbf{73.33\%\\}
    99 respondents use ChatGPT to explore alternative (better) ways to develop code or implement specific features. This indicates that ChatGPT plays a crucial role in offering creative insights and ideas to developers during the software development lifecycle.
    \item {\textit{\textbf{Library Selection:}}~\color{black!90}\percentagebar{24pt}} \textbf{46.67\%\\}
    ChatGPT is considered by 63 participants to discover appropriate libraries and frameworks, for their development projects.
    \item {\textit{\textbf{Others:}}~\color{black!90}\percentagebar{2pt}} \textbf{2\%\\}
    Some developers (2\%) also mentioned additional use cases for ChatGPT, such as comparing code, using it for non-professional coding needs on websites, performing unit tests, etc. 

\end{itemize}

\nd\circled{Q4} Next we asked the participants about their reaction/experience of using ChatGPT for different development tasks. The participants expressed both positive and negative feelings as follows. 

\begin{itemize}[leftmargin=9pt,nolistsep]
    \setlength\itemsep{0.2em}
    \item \textit{\textbf{\small Exciting:}}~\color{black!90}\percentagebar{34pt} \textbf{68.15\%},
    \textit{\textbf{\small Tech Marvel:}}~\color{black!90}\percentagebar{20pt} \textbf{42.22\%\\}
    92 respondents reported being excited by ChatGPT's capabilities and potential to assist them in various tasks while 57 of them remark it as a technological marvel.  
    \item {\textit{\textbf{\small Unreliable:}}~\color{black!90}\percentagebar{15pt}} \textbf{34\%},
    {\textit{\textbf{\small Over-hyped:}}~\color{black!90}\percentagebar{6pt}} \textbf{19.26\%},
    {\textit{\textbf{\small Suspicious:}}~\color{black!90}\percentagebar{3pt}} \textbf{9.63\%\\}
    46 participants found their experience with ChatGPT to be unreliable. They encountered instances where the tool provided inaccurate or inconsistent responses. 26 respondents feel considered ChatGPT overhyped and the tool's capabilities as exaggerated. 
\end{itemize}

\begin{tcolorbox}[
       left=0pt, right=0pt, top=0pt, bottom=0pt, colback=white, after=\ignorespacesafterend\par\noindent]
\textbf{Summary of RQ1.} \rev{The majority of developers (98.52\%) used ChatGPT for various purposes. From code generation (60\%) and analysis (52.59\%) to problem-solving (86.67\%), exploring coding alternatives (73.33\%), and libraries (46.67\%), ChatGPT served as a versatile tool to them. Other common usages include information retrieval (79.26\%), content creation (46.67\%), etc.}
\end{tcolorbox}

\subsection{Concerns about ChatGPT Responses (RQ2)}
\circled{Q5} We asked participants to indicate the extent of their reliance on ChatGPT contents, revealing a spectrum of responses ranging from complete trust to cautious validation with other sources. Most participants are found to be doubtful of ChatGPT responses.

\begin{itemize}[leftmargin=9pt,nolistsep]
    \setlength\itemsep{0em}
    \item {\textit{\textbf{Not Reliable At All:}}~\color{black!90}\percentagebar{4pt}} \textbf{5.19\%\\}
    7 participants indicated that they do not consider the ChatGPT responses as reliable at all. They always cross-check the information with other sources for validation. 
    \item \textit{\textbf{Somewhat Reliable:}} \textit{\textbf{Need Validation}}~\color{black!90}\percentagebar{27pt} \textbf{54.81\%,\\} \textit{\textbf{Need Augmentation}}~\color{black!90}\percentagebar{17pt} \textbf{35.56\%\\}
    The majority (74 participants) stated that while ChatGPT provides them with initial ideas, they rely on other sources for validation. Moreover, 48 participants expressed that they subsequently search for additional knowledge from other sources. Only one participant expressed reliance on the correctness of responses. 
    \item  {\textit{\textbf{Others: }}~\color{black!90}\percentagebar{2pt}} \textbf{3.70\%\\}
    A few (5) participants put emphasis on proper prompting to assess inaccuracy, i.e., asking questions to ChatGPT more precisely. 
\end{itemize}

\nd\circled{Q6} We asked participants to discuss the pros and cons of ChatGPT usage. Specifically, we asked for their opinion on library selection and comparison. Out of the 135 participants, 72 provided responses to this open-ended question. The results are summarized below.

\nd\underline{\textbf{Pros of Using ChatGPT for Library Selection}}

\begin{itemize}[leftmargin=10pt,nolistsep]
    \item \textit{\textbf{Efficient Access to Information:}} 22 participants mentioned that ChatGPT provides quick and efficient access to information which allows them to receive library suggestions or comparisons.

    \item \textit{\textbf{Initial Idea Generation:}} 10 respondents found ChatGPT helpful for generating initial ideas or suggestions regarding software libraries, serving as a starting point for further research.

    \item \textit{\textbf{Personalized Recommendations:}} 3 participants appreciated that ChatGPT could offer personalized recommendations based on specific requirements or preferences.
    
    \item \textit{\textbf{Time-Saving:}} 10 respondents highlighted that ChatGPT's ability to provide library suggestions or comparisons quickly saved them time during the decision-making process.
\end{itemize}

\nd\underline{\textbf{Cons of Using ChatGPT for Library Selection}}
\begin{itemize}[leftmargin=10pt,nolistsep]
    \item \textit{\textbf{Lack of Up-to-dateness:}} 12 participants expressed concerns about ChatGPT's limitations compared to human experts. They mentioned that it might lack up-to-date knowledge, real-world experiences, or insights about specific libraries.

    \item \textit{\textbf{Contextual Understanding Challenges:}} 11 respondents noted that ChatGPT might struggle to understand nuanced or complex questions related to software libraries, leading to potentially inaccurate or irrelevant recommendations.
    
    \item \textit{\textbf{Reliability Concerns:}} 26 participants pointed out that ChatGPT's responses were not always accurate or reliable, suggesting the need to cross-check with other sources for validation.
    
    \item \textbf{\textit{Dependence on Prompt:}} 4 respondents mentioned that the quality of ChatGPT's responses relied on the input (prompt) provided by users, leading to biased/misguided recommendations.
    
    \item \textit{\textbf{Not Sufficient for Decision-Making:}} 7 participants emphasized that ChatGPT might be inaccurate  to be solely relied upon for complex decision-making tasks.

    \item \textit{\textbf{Bias of Training Data:}} A few participants indicated the possible bias that might exist in ChatGPT's response. 

\end{itemize}

\begin{tcolorbox}[
       left=0pt, right=0pt, top=0pt, bottom=0pt, colback=white, after=\ignorespacesafterend\par\noindent]
\textbf{Summary of RQ2.} There are concerns about ChatGPT reliability, with some finding it unreliable (34\%) and over-hyped (19.26\%). Only a small portion (0.74\%) of our survey participants fully rely on the responses of ChatGPT (i.e., without double-checking). 
\end{tcolorbox}

\subsection{Verification of ChatGPT Responses (RQ3)}
\nd\circled{Q7} Participants were presented with our conversation with ChatGPT where we asked  ChatGPT about a good natural language processing (NLP) library in Python. ChatGPT suggested\texttt{spaCy}, \texttt{NLTK}, and \texttt{Gensim} as notable options. Our conversation with ChatGPT is as follows. 
{\small
\begin{tcolorbox}[enhanced,
      boxsep=0pt,top=0pt,bottom=0pt,leftupper=0pt,rightupper=0pt,
      toptitle=1mm,bottomtitle=1mm,lefttitle=5mm,righttitle=5mm,
      colbacktitle=gray!20,coltitle=black,title=\textbf{Interaction with ChatGPT for Survey Question \#7},
      colback=white!10,clip upper]%
    \begin{tcbitemize}[raster columns=1,raster equal skip=0pt,
        sharp corners,boxrule=0pt,
        raster odd row/.style={empty},
        raster even row/.style={tile,colback=lightgray!20}]%
      \tcbitem
      \usericon~\textbf{Prompt:} What is a good library in Python for NLP? 
      \tcbitem
      \chatgpticon~\textbf{Response:} spaCy is considered a top library for natural language processing (NLP) in Python. It provides a wide range of NLP functionalities, efficient processing, support for multiple languages, and easy integration with deep learning frameworks. Other notable options include NLTK and Gensim. 
    \end{tcbitemize}
\end{tcolorbox}
} 

Participants were then asked whether they would rely on this response or not, and how they would verify it further. While a tiny portion (2.22\%) of participants were ready to accept ChatGPT response without further verification, most of them preferred to seek additional information and verify the response as follows.

\begin{itemize}[leftmargin=9pt,nolistsep]
    \setlength\itemsep{0em}
    \item {\textit{\textbf{Further Search in Common Medium:}}~\color{black!90}\percentagebar{37pt}} \textbf{73.33\%\\}
    The majority of participants (99 respondents) preferred to supplement ChatGPT's response by conducting further research on their own (e.g., by consulting Google or Stack Overflow). 
    \item \textit{\textbf{Further Inquiry to ChatGPT:\\}} \textit{\textbf{About spaCy}}~\color{black!90}\percentagebar{34pt} \textbf{68.15\%,} \textit{\textbf{NLTK\&Gensim}}~\color{black!90}\percentagebar{22pt} \textbf{45.19\%\\}
    92 participants stated that they would ask ChatGPT more questions to further assess whether a suggested library (e.g., \texttt{spaCy}) fits their specific needs. 
    \item {\textit{\textbf{Others: }}~\color{black!90}\percentagebar{4pt}} \textbf{4.44\%\\}
    A few respondents mentioned some additional ways to verify the response such as seeing the package/library or GitHub repository, reading the documentation, etc.
\end{itemize}

\nd\circled{Q8} Participants were presented with a specific situation where ChaGPT is asked for more details about \texttt{spaCy}'s API and documentation. They were then asked about their level of reliance on this detail. The interaction with ChatGPT is as follows.

{\small
\begin{tcolorbox}[enhanced,
      boxsep=0pt,top=0pt,bottom=0pt,leftupper=0pt,rightupper=0pt,
      toptitle=1mm,bottomtitle=1mm,lefttitle=5mm,righttitle=5mm,
      colbacktitle=gray!20,coltitle=black,title=\textbf{Interaction with ChatGPT for Survey Question \#8},
      colback=white!10,clip upper]%
    \begin{tcbitemize}[raster columns=1,raster equal skip=0pt,
        sharp corners,boxrule=0pt,
        raster odd row/.style={empty},
        raster even row/.style={tile,colback=lightgray!20}]%
      \tcbitem
      \usericon~\textbf{Prompt:} How easy it is to use spaCy? 
      \tcbitem
      \chatgpticon~\textbf{Response:} spaCy is known for its user-friendly design, making it easy to use for natural language processing (NLP). With a straightforward API, extensive documentation, and many pre-trained models, spaCy simplifies the process of developing applications, making it accessible to beginners and advanced users. 
    \end{tcbitemize}
    
\end{tcolorbox}
}
After seeing this additional confirmatory prompt, developers have mixed expressions about the reliability of the previous response (i.e., \texttt{spaCy} to be a top NLP library) in survey question \textit{Q7}. Some of them have shifted towards accepting the response while others still wanted further verification.

\begin{itemize}[leftmargin=9pt,nolistsep]
    \setlength\itemsep{0em}
    \item {\textit{\textbf{Rely on Confirmatory Details:}}~\color{black!90}\percentagebar{28pt}} \textbf{55.56\%\\}
    75 respondents demonstrated high confidence in the new details and indicated that they would now rely on the information provided. They expressed their intention to jump into the official documentation of \texttt{spaCy} to learn more.
    \item {\textit{\textbf{Still Search in Common Medium:}}~\color{black!90}\percentagebar{25pt}} \textbf{51.11\%\\}
    69 participants still wanted to look for additional opinions and insights from sources like Google and Stack Overflow.
    \item \textit{\textbf{Further Inquiry to ChatGPT:}} \textit{\textbf{About spaCy}}~\color{black!90}\percentagebar{24pt} \textbf{49.63\%\\}
    67 respondents sought further clarification by asking the ChatGPT for sample APIs or methods related to \texttt{spaCy}. They wanted more specifications for better understanding and verification.
    
\end{itemize}

\nd\circled{Q9} To ensure ChatGPT's reliability in SE real-world situations, we need to assess its ability to- i) interpret specific scenarios and contexts accurately; ii) generate appropriate information based on that. Hence, we provided ChatGPT with a specific question-answer set (about \texttt{spaCy}) from Stack Overflow and asked it to interpret whether the library is easy to use solely based on the provided information. Our interaction with ChatGPT is as follows.

{\small
\begin{tcolorbox}[enhanced,
      boxsep=0pt,top=0pt,bottom=0pt,leftupper=0pt,rightupper=0pt,
      toptitle=1mm,bottomtitle=1mm,lefttitle=5mm,righttitle=5mm,
      colbacktitle=gray!20,coltitle=black,title=\textbf{Interaction with ChatGPT for Survey Question \#9},
      colback=white!10,clip upper]%
    \begin{tcbitemize}[raster columns=1,raster equal skip=0pt,
        sharp corners,boxrule=0pt,
        raster odd row/.style={empty},
        raster even row/.style={tile,colback=lightgray!20}]%
      \tcbitem
      \usericon~\textbf{Prompt:} How easy it is to use the library spaCy strictly based on the following conversation (\textit{Question, Answer}).
      
      \textit{Question: I'm processing 40,000 abstracts with spacy nlp.pipe and its taking 8 mins. Is there a way to speed this up further? I've also disabled ner.}
      
      \textit{Answer: Try tuning batch\_size and n\_process params: Note as well, by joining on "" "" you may have some surprises, as spaCy's splitting rules are a bit more complex than that.}
      \tcbitem
      \chatgpticon~\textbf{Response:} spaCy is easy to use, but optimizing performance may involve tuning batch size and process count parameters.
    \end{tcbitemize}
\end{tcolorbox}
}

We showed this interaction with ChatGPT to the participants and asked them whether they agree with ChatGPT's interpretation about the usability of Spacy. Most of them remain unconvinced.

\begin{itemize}[leftmargin=9pt,nolistsep]
    \setlength\itemsep{0em}
    \item {\textit{\textbf{Unsure of ChatGPT's Interpretation:}}~\color{black!90}\percentagebar{27pt}} \textbf{54.07\%\\}
    73 respondents were unsure about the accuracy of ChatGPT's interpretation. They neither fully agreed nor disagreed.
    \item {\textit{\textbf{Disagree with ChatGPT's Interpretation:}}~\color{black!90}\percentagebar{18pt}} \textbf{34.81\%\\}
    47 respondents disagreed with ChatGPT's response, suggesting that the conversation actually discussed the complexity of spaCy and not its ease of use. 
    \item {\textit{\textbf{Agree with ChatGPT's Interpretation:}}~\color{black!90}\percentagebar{5pt}} \textbf{11.11\%\\}
    15 respondents agreed with ChatGPT's interpretation, stating that the conversation somehow hints that spaCy is easy to use.
\end{itemize}

\nd\circled{Q10} Finally, participants were asked to suggest ways to improve the reliability of ChatGPT responses. The participants' responses can be grouped into the following key ways.

\begin{itemize}[leftmargin=9pt,nolistsep]
    \setlength\itemsep{0em}
    \item {\textit{\textbf{Showing References:}}~\color{black!90}\percentagebar{40pt}} \textbf{80.74\%\\}
    109 participants suggested that showing (online) references supporting the answer would improve the reliability of ChatGPT responses. One participant proposed showing diagrams and resource links (e.g., YouTube tutorials) as supplementary materials.
    \item {\textit{\textbf{Asking Challenging Questions:}}~\color{black!80}\percentagebar{22pt}} \textbf{44.44\%\\}
    60 participants mentioned that repeatedly asking challenging questions to ChatGPT could help us with appropriate direction.
    \item {\textit{\textbf{Better Prompt Engineering:}}~\color{black!90}\percentagebar{3pt}} \textbf{3.70\%\\}
    5 participants highlighted the importance of improving prompt engineering to make ChatGPT responses more reliable.
    \item {\textit{\textbf{User Feedback:}}~\color{black!90}\percentagebar{2pt}} \textbf{1.48\%\\}
    2 participants emphasized the significance of user feedback and learning to improve the ChatGPT's response quality. 
    \item {\textit{\textbf{Other Chatbots' Feedback:}}~\color{black!90}\percentagebar{1pt}} \textbf{0.74\%\\}
    1 participant came up with the idea of verifying ChatGPT responses from other similar chatbots (such as Google BARD). 
\end{itemize}

\begin{tcolorbox}[
       left=0pt, right=0pt, top=0pt, bottom=0pt, colback=white, after=\ignorespacesafterend\par\noindent]
\textbf{Summary of RQ3.} The majority (54.81\%) consider ChatGPT responses somewhat reliable and seek further validation. When verifying them, most developers (73.33\%) conduct further searches in common mediums like Google or Stack Overflow to gather additional information. Many participants (68.15\%) would ask ChatGPT more questions to assess its responses further. 
\end{tcolorbox}

\section{CID: An Automatic \underline{C}hatGPT \underline{I}ncorrectness \underline{D}etector}
\label{sec:cid}
Our survey participants wished to perceive an LLM as a remarkably close approximation of human capabilities. A desirable property of a good LLM is consistency: the ability to make consistent decisions in semantically equivalent
contexts. However, it is true that humans are also prone to inconsistency, especially when they are lying or bluffing. There is a general tendency among practitioners (e.g., police officers) to perceive consistent statements as truthful, and inconsistent statements as deceptive \cite{deeb2018police}. \jRevision{In fact, studies in criminal justice find that increasing the cognitive load on the suspect with unanticipated (e.g., convoluted) questions is effective in outsmarting/detecting liars \cite{vrij2009outsmarting, vrij2011outsmarting, lancaster2013sorting}. A recent study by Dhuliawala et al. demonstrates the effectiveness of similar approaches on LLMs \cite{dhuliawala2023chain}. They proposed the Chain-of-Verification (CoVe) method, which challenges the model to verify its initial responses with independently generated questions to enhance the reliability of the model's output.} Therefore, our CID (ChatGPT Incorrectness Detector) tool uses iterative prompting to capture ChatGPT's inconsistency in a similar fashion to an actual Crime Investigation Department (CID). 

\begin{figure}[h]
    \centering
    \includegraphics[scale=.68]{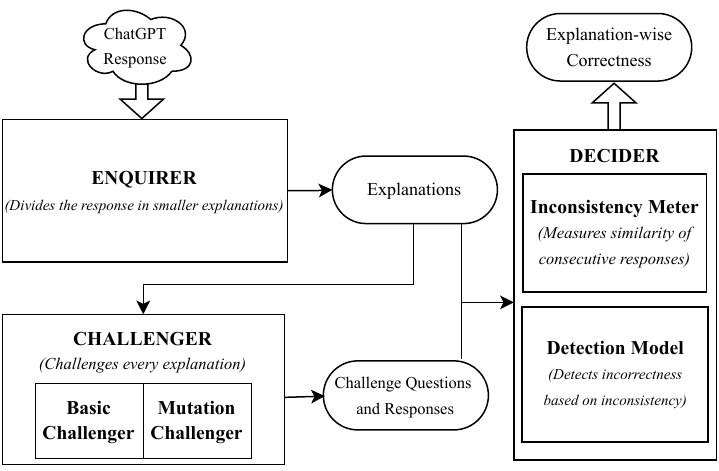}
    \caption{Overview of CID Tool.}
    \label{fig:overview_CID}
\end{figure}

In a real case, an interrogator might: \textbf{1) Enquire} the suspect about the particular case, \textbf{2) Challenge} the suspect to reveal any inconsistency, \textbf{3) Decide} on the responses of the suspect. Likewise, CID comprises three key components: ENQUIRER, CHALLENGER, and DECIDER. The overview of CID is shown in Figure \ref{fig:overview_CID}. The ENQUIRER seeks the initial reasoning behind ChatGPT's responses, while the CHALLENGER poses a series of questions to induce potential inconsistencies. The DECIDER employs machine learning (ML) techniques and similarity measures to determine the correctness of ChatGPT's responses based on the detected inconsistency. We discuss each component in detail below.

\subsection{ENQUIRER}
Suppose, we ask a question (base-question, $Q_B$) and want to evaluate the correctness of its response (base-response, $R_B$). The ENQUIRER targets to obtain ChatGPT's initial reasoning behind the base-response that can be useful to reveal any inconsistency in the next steps of interrogation. When ChatGPT responds to any question, it typically provides multiple pieces of information, which may lead to partial correctness, meaning that some parts of the response are accurate while others are not. To address this issue, the ENQUIRER asks ChatGPT to provide separate explanations ($E_i$) for each piece of information that exists in the base-response. The ENQUIRER performs this by using the following prompt.

{\small
\begin{tcolorbox}[enhanced,
      boxsep=0pt,top=0pt,bottom=0pt,leftupper=0pt,rightupper=0pt,
      toptitle=1mm,bottomtitle=1mm,lefttitle=5mm,righttitle=5mm,
      colbacktitle=gray!20,coltitle=black,title=\textbf{Enquiring ChatGPT},
      colback=white!10,clip upper]%
    \begin{tcbitemize}[raster columns=1,raster equal skip=0pt,
        sharp corners,boxrule=0pt,
        raster odd row/.style={empty},
        raster even row/.style={tile,colback=lightgray!20}]%
      \tcbitem
      \faTools~\textbf{ENQUIRER:} Justify your answer. If the answer has multiple pieces of information, provide separate reasoning for each of them.
    \end{tcbitemize}
\end{tcolorbox}
}

\subsection{CHALLENGER} For every piece of explanation ($E_i$) obtained in the ENQUIRY phase, We further ask a set of challenge-questions ($Q_C$) to ChatGPT through our CHALLENGER component and record chatGPT's responses (challenge-responses, $R_C$) to them. The basic intuition is that for a piece of information given by an LLM that is possibly incorrect, the LLM will be (more) inconsistent in such consecutive prompts. The Challenger component works in four steps. In the first step, it generates a dynamic challenge question for a previous explanation; in the second step, it asks the generated question to ChatGPT; in the third step, it mutates the challenge and finally in the fourth step, it asks the mutated challenge again to ChatGPT. We describe the basic and mutation challenges below.

\subsubsection{Basic Challenger} We first ask three basic challenge questions to ChatGPT:  {\textit{Why?}}, {\textit{How?}}, {\textit{Really?}} for each explanation ($E_i$) of its base-response ($R_B$). \jRevision{To automate the questioning process, the basic challenger leverages a separate LLM (completely unrelated to the original ChatGPT instance that is being evaluated by the CID tool). To replicate a separate LLM, we used ChatGPT with a new separate session. The motive for using a separate session of ChatGPT is to discard the memory of the previous conversation performed (with ChatGPT) before generating the challenge questions (i.e., during the Enquiry phase). This is done to ensure the elimination of potential bias that could be induced from the ENQUIRER memory.} We generate the basic challenge questions as follows.

{\small
\begin{tcolorbox}[enhanced,
      boxsep=0pt,top=0pt,bottom=0pt,leftupper=0pt,rightupper=0pt,
      toptitle=1mm,bottomtitle=1mm,lefttitle=5mm,righttitle=5mm,
      colbacktitle=gray!20,coltitle=black,title=\textbf{Generating Basic Challenge Questions},
      colback=white!10,clip upper]%
    \begin{tcbitemize}[raster columns=1,raster equal skip=0pt,
        sharp corners,boxrule=0pt,
        raster odd row/.style={empty},
        raster even row/.style={tile,colback=lightgray!20}]%
      \tcbitem
      \faTools~\textbf{BASIC-CHALLENGER:} Generate a question that starts with [Why/How/Really] to challenge the following [EXPLANATION].
    \end{tcbitemize}
\end{tcolorbox}
}

\subsubsection{Mutation Challenger} Returning to our analogy between human deception and LLMs' errors: (human) liars can often be as consistent as truth-tellers since they prepare for the interview by anticipating questions and rehearsing responses to them. However, the interviewer may make the interview more cognitively demanding for liars which disrupts liars’ preparations and makes it difficult for them to provide spontaneous responses. Similarly, our confirmatory questions, if simple, might not pose enough challenge for the LLM which means the LLM could answer them consistently even for a wrong base-response/explanation. Hence, we use Mutation Challenger to further modify or complicate the challenge questions so that they can be more difficult for ChatGPT. Therefore, the Mutation Challenger mutates the basic challenge questions to create mutation challenge questions ($Q_{CB}\rightarrow Q_{CM}$). 

\begin{figure}
    \centering
    \includegraphics[scale=.65]{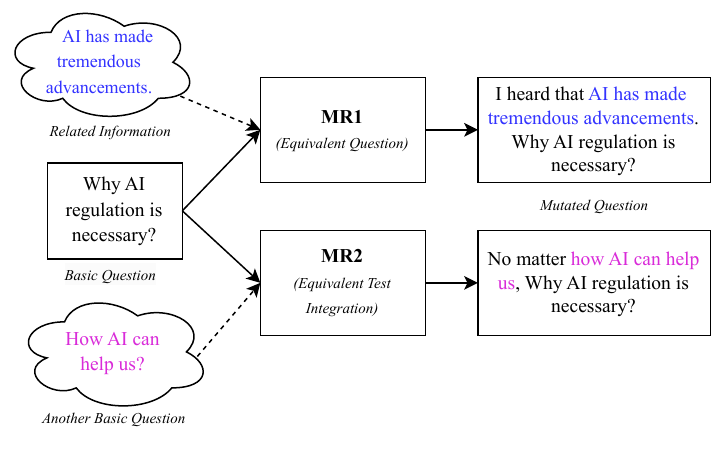}
    \caption{Metamorphic Relations (MRs) used in the Mutation Challenger to mutate questions.}
    \label{fig:MR_daigram}
    \vspace{-3mm}
\end{figure}

\rev{This approach benefits from existing techniques for evaluating question-answer (QA) language models \cite{zhang2020machine, clark2019boolq, hirschman2001natural, ribeiro2020beyond}. Multiple studies proposed reliability testing techniques for the QA software \cite{ribeiro2020beyond, eger2020hero, jia2017adversarial, longpre2021entity, chen2021testing, shen2022natural}. Among them, Shen et al. \cite{shen2022natural} outperformed previous state-of-the-art approaches by discovering 23\% more bugs or inconsistencies in the target answer-generating models and the Challenger component adopted their question mutation technique for generating challenging questions.
}
For this, our mutation challenger employs the sentence-level metamorphic testing technique, QAQA, proposed by \cite{shen2022natural}. It inserts a redundant sentence as a clause to the original (basic challenge) question to generate the mutated question which puts a larger cognitive load on ChatGPT and challenges it further to reveal inconsistency with the responses ($R_{CM}$). The redundant sentence provides an additional fact that is combined with the original question without altering its semantics or the fact itself. Depending on the source of the redundant sentence, the mutation challenger applies two types of metamorphic relation (MR): \textbf{Equivalent Question (MR1)} and \textbf{Equivalent Test Integration (MR2)}. While MR1 chooses the redundant information from a pre-defined knowledge base, MR2 extracts it from other basic challenge questions from the input list. In both cases, the information (or the question) is selected based on their cosine similarity with the original basic challenge question. We show examples of both MRs in Figure \ref{fig:MR_daigram}. Different subordinate clauses (such as \textit{`I heard that'}, \textit{`No matter what'}, \textit{`I do not care'}, etc.) are used to combine the redundant information with the original question.


\subsection{DECIDER}
\label{sec:decider}
DECIDER makes the final call about the (in)correctness of a given response. It has two modules.

\begin{itemize}[leftmargin=9pt,nolistsep]
    \item Inconsistency Meter: This part of DECIDER evaluates ChatGPT's inconsistency during the ENQUIRY and CHALLENGE phases using standard similarity metrics (e.g., Cosine Similarity).
    \item Detection Model: DECIDER utilizes an ML model (e.g., Support Vector Machine) to make the final decision on the correctness of ChatGPT's response. This model is trained to learn the relation between ChatGPT's (in)correctness and (in)consistency from a labelled dataset. By doing so, DECIDER can effectively detect whether a response is correct or not based on the level of consistency.
\end{itemize}

\subsubsection{Dataset Creation.} To train the DECIDER (specifically, its detection model), we need a labeled dataset. We generate the dataset by interacting with ChatGPT and posing various questions to it. Each question should have a known correct answer that serves as the ground truth. For each question in the dataset, we interact with ChatGPT and record its base response to the given question. This base response will serve as the initial answer to be evaluated. Then we use our ENQUIRER to split each base response into multiple explanations ($E_i$). Finally, these explanations are manually labeled as correct/incorrect by human annotators. In Section \ref{sec:benchmark}, we discuss how we created such a benchmark.

\subsubsection{Inconsistency Meter and Detection Model.} Once we have a labeled dataset with explanations ($E_i$) labeled as correct or incorrect, we proceed with the feature extraction process. We compute standard similarity scores among ChatGPT responses generated in the ENQUIRY and CHALLENGE phases; use them as features for our tool. We then train an ML model that learns the relation between ChatGPT's incorrectness (i.e., labels) and inconsistency (features). We consider 24 features from four categories.      

\begin{itemize}[leftmargin=9pt,nolistsep]
    \item {Explanation-Response (\textit{E\textsubscript{i}-R\textsubscript{C}}) Similarity:} This category includes similarity scores calculated between the explanation and each of the challenge-responses. Hence, we get six similarity scores here (three for basic and three for mutation challenges). This allows our system to detect potential inconsistencies in the responses with respect to the base explanation.

    \item {Response-Response (\textit{R\textsubscript{C}-R\textsubscript{C}}) Similarity:} The similarity among every pair of challenge responses is considered in this category. As the challenge responses point to the same piece of explanation, the system aims to identify potential discrepancies or inconsistencies between them. Higher similarity scores between challenge responses could indicate a consistent and well-supported base response/explanation, while lower similarity scores may suggest conflicting information or possible incorrectness. For pair-wise similarity among 3 basic challenge responses, we get $3\choose2$ = 3 features. Similarly, we get another 3 features for the 3 mutation challenge responses (i.e., total  six \textit{R\textsubscript{C}-R\textsubscript{C}} features).

    \item {Question-Response (\textit{Q\textsubscript{C}-R\textsubscript{C}}) Similarity:} This category involves pair-wise similarity calculations between each challenge-question and corresponding challenge-response. The purpose is to see if the responses are relevant to the challenge questions. We again get six similar features from this category. 
    
    \item {Question-Question (\textit{Q\textsubscript{C}-Q\textsubscript{C}}) Similarity:} While we expect $R_C-R_C$ pairs to be consistent with each other, and consider their similarity scores as features, their inconsistency could be associated to the inconsistency/dissimilarity between the corresponding challenge questions ($Q_C$). This category aims to inform the model of pair-wise (di)similarities of the challenge questions resulting in six $Q_C-Q_C$ similarity features.
    
\end{itemize}

\section{Evaluation of CID}\label{sec:evaluation}
Given that CID can be applicable for any computing tasks that
require the usage of ChatGPT textual responses, within the short space
of this paper we focused on evaluating CID for a given SE task: software library selection. We picked this task because our survey responses showed SE practitioners want to use ChatGPT during their selection and reuse of a software library. 
\rev{Moreover, various research papers in SE literature reported that software library selection can be a non-trivial task for developers \cite{wasserman2017osspal, li2022exploring, larios2020selecting, huang2018tell, wang2020difftech, wang2021difftech, uddin2017automatic, uddin2017opiner, de2018library, de2018empirical, el2020libcomp, yan2022concept, liu2021api, uddin2019understanding}.}
We thus evaluated CID using a benchmark study where the benchmark has information about correct and incorrect ChatGPT responses regarding software library usage. Using this benchmark dataset, we answer two RQs:
\begin{enumerate}[label=\textbf{RQ\arabic{*}.}, start=4,leftmargin=25pt]
    \item How accurate is CID in detecting incorrect responses? 
    \item How do the base and mutation challenge prompts impact the performance?
\end{enumerate}

\subsection{Benchmark Study Setup}\label{sec:benchmark}
\subsubsection{Benchmark Dataset Collection} We first selected 100 Stack Overflow (SO) posts regarding which we will ask the questions.  The posts were related to a few popular text processing libraries (\href{https://spacy.io}{spaCy}, \href{https://www.nltk.org}{NLTK}, and \href{https://github.com/google/gson}{GSON}). 
From the posts, we collected the titles, questions, and accepted answers. Such accepted answers could provide ground truth to assess ChatGPT responses.

\rev{
We followed a structured approach to collect the 100 posts using the search interface of SO \cite{website:so-search}. In the search page, we provided two criteria: the post must have an accepted answer and the post must have tags of one of the popular text-processing libraries (GSON, NLTK, or spaCy). We sorted the search results so that the posts with the highest scores come first (the SO community sets the score). Then we manually analyzed each post based on two more criteria: the post should discuss any of seven technical aspects (such as Active Maintenance, Documentation, Ease of use, Feature, Performance, Security, and Stability) that developers use for evaluating libraries \cite{larios2020selecting}. We manually analyzed nearly 500 posts and stopped further analysis after collecting 100 posts in total since the analysis was an extensively labor-intensive process. The reasoning of the selection criteria and the aspect-specific post count details is provided in online the appendix \cite{website:replication-package}.
}

\begin{table}[t]
    \centering
    \caption{Base question template for the SO posts}
    \label{tab:base_prompt_so}
    \begin{tabular}{ll}
        \toprule
        \textbf{Factor} & \textbf{Base Question ($Q_B$)}  \\ \midrule
        Active Maint. & How actively the library is maintained \\ 
        Documentation & How is the documentation of the library \\ 
        Ease of use & How easy it is to use the library \\ 
        Feature & How well does this library support \textit{[x]} feature \\ 
        Performance & How is the performance of the library \\ 
        Security & How is the security of the library \\ 
        Stability & How stable or well tested is the library \\
        \bottomrule
    \end{tabular}
\vspace{-4mm}
\end{table}

\subsubsection{Base Question Generation} 
\rev{
We used Chat API interface of ChatGPT model version \href{https://platform.openai.com/docs/models/overview}{gpt-3.5-turbo-0301}. }
We created a question template for each of the factors.
Table \ref{tab:base_prompt_so} shows the list of base questions we used for different SO posts. 
All the base questions ask for an opinion from ChatGPT regarding the library factor or feature. Instead of giving generic responses to the base questions, we wanted ChatGPT to provide the response only based on the related SO data. Hence, when we created the basic question for ChatGPT, we also added the SO post content (SO title, SO question, SO answer) as a context in the question. The prompt has the following format: 
\textit{``Respond in less than 200 words "+Base Question ($Q_B$) + " strictly based on the following conversation (question, answer): + Context (C)''.} For example, a query for ease of use factor would be: 

{\small
\begin{tcolorbox}[enhanced,
      boxsep=0pt,top=0pt,bottom=0pt,leftupper=0pt,rightupper=0pt,
      toptitle=1mm,bottomtitle=1mm,lefttitle=5mm,righttitle=5mm,
      colbacktitle=gray!20,coltitle=black,title=\textbf{Enquiring ChatGPT},
      colback=white!10,clip upper]%
    \begin{tcbitemize}[raster columns=1,raster equal skip=0pt,
        sharp corners,boxrule=0pt,
        raster odd row/.style={empty},
        raster even row/.style={tile,colback=lightgray!20}]%
      \tcbitem
      \faTools~\textbf{Base Question:} Respond in less than 200 words How easy it is to use the library strictly based on the following conversation (question, answer):
      
      Question: <SO Question>
      Answer: <SO Answer>
    \end{tcbitemize}
\end{tcolorbox}
}

\subsubsection{Explanation Generation from ChatGPT}
After the base response is generated from ChatGPT, we use the ENQUIRER component to generate further explanation of the base response. The base response received from ChatGPT is an aggregated opinion and some parts of the opinion can be correct and some parts can be incorrect. 

\rev{
To identify the exact correctness, we have to dissect the opinion into multiple reasons. The first part of the following example shows the format of an exploratory question asked to ChatGPT. It contains a prompt to explain the previous base response and also provides the previous base question and answer as a reference. The second part of the example shows the response statements generated by ChatGPT. For simplicity, we omitted the JSON formatting from the response and only provided the statements. We also marked the correctness of those statements with green/red icons (however, the correctness is detected by the last component Detector). 
}

{\small
\begin{tcolorbox}[enhanced,
      boxsep=0pt,top=0pt,bottom=0pt,leftupper=0pt,rightupper=0pt,
      toptitle=1mm,bottomtitle=1mm,lefttitle=5mm,righttitle=5mm,
      colbacktitle=gray!20,coltitle=black,title=\textbf{Enquiring ChatGPT},
      colback=white!10,clip upper]%
    \begin{tcbitemize}[raster columns=1,raster equal skip=0pt,
        sharp corners,boxrule=0pt,
        raster odd row/.style={empty},
        raster even row/.style={tile,colback=lightgray!20}]%
      \tcbitem
      \faTools~\textbf{\rev{Explanatory Question:}} 
      Provide explanation for the answer. Each reason separately with title in 4 words and the explanation in 50 words. Respond strictly in JSON [{"title":x, "explanation":y}] strictly based on the previous conversation (question, answer).
      
      Question: <SO Question>
      Answer: <SO Answer>
      
      <Base Question>
      <Base Reponse>
      \tcbitem
      \chatgpticon~\textbf{Response:} 

        \textcolor{green}{\cmark}~{A bug affects Spacy 2.3.0 and its tagger training functionality.}

        \textcolor{green}{\cmark}~{spaCy dev team maintains and updates the library regularly.}
        
        \textcolor{red}{\xmark}~{spaCy dev team is responsive and attentive to user feedback.}

    \end{tcbitemize}
\end{tcolorbox}
}

\subsubsection{Explanation Labeling} After asking this enquiring question, ChatGPT responded with 3/4 explanations for each of the questions. Thus, for 100 posts, we found 341 explanations (in total) from ChatGPT. According to our labeling based on the original SO post, 276 explanations (81\%) of them were correct.

\subsection{Incorrectness Detection Performance (RQ4)}
\subsubsection{Approach} First, we trained and tested three supervised ML models (logistic regression (LR), random forest (RF), and support vector machine (SVM)) on our benchmark using a 10-fold cross-validation. We report the performance of the ML models on their incorrectness detection using four metrics: precision (P), recall (R), F1-score (F1), and accuracy (A). Second, we manually assess the misclassifications of our best-performing model.

\subsubsection{Results}
\label{sec:results}
\begin{figure}
    \centering
    \includegraphics[scale=.40]{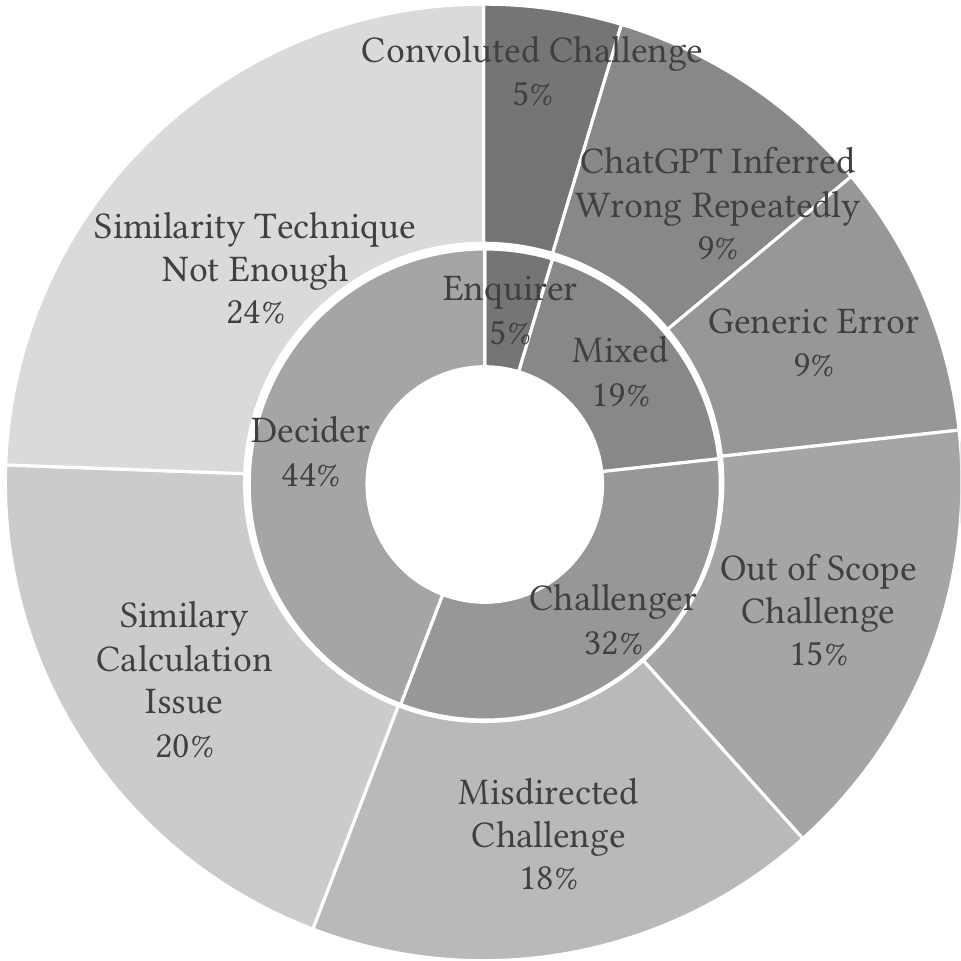}
    \caption{Sources and categories of misclassification.}
    \label{fig:misclassification}
\end{figure}

The evaluation results are shown in Table \ref{tab:model_accuracy}.
\begin{table}[]
    \centering
    \caption{ML model performance to detect ChatGPT incorrectness. P = Precision, R = Recall, A = Accuracy. F1 = F1-score}
    \label{tab:model_accuracy}

    \begin{tabular}{lrrrr}
    \toprule
     \textbf{Model} & \textbf{P} & \textbf{R} & \textbf{A} & \textbf{F1} \\ \midrule
    Logistic Regression (LR) & 0.74 & 0.65 & 0.65 & 0.68 \\ 
    Random Forest (RF) & 0.73 & 0.65 & 0.65 & 0.68 \\ 
    Support Vector Machine (SVM) & 0.74 & 0.75 & 0.75 & \textbf{0.74} \\ 
        
    \bottomrule
    \end{tabular}
\end{table}
SVM outperformed LR and RF models with an F1-score of 0.74, and accuracy 0.75 which means the decider could identify the correctness of 259 out of 341 ChatGPT responses. We further reviewed all 86 misclassifications from SVM model and identified 7 misclassification categories originating from 4 different sources (Figure \ref{fig:misclassification}). 

\nd\ul{\textbf{Error Source - Decider (44\% of Errors).}} 
24\% errors occur because the current \textbf{similarity technique is insufficient} to detect incorrect responses. The similarity score can be high even for an incorrect response when all challenge responses admit the mistake unanimously. An additional layer can be added to the decider 
to identify and filter out such unanimous confession of the challenge responses. Another category is the \textbf{similarity calculation issue} (21\%) when the sentence-based transformers struggle to accurately detect high similarity among larger text responses from ChatGPT.

\nd\ul{\textbf{Error Source - Challenger (32\%).}} 
18\% errors are due to \textbf{misdirected challenges} where the challenger generates a question that is unrelated to the actual logical reasoning of the explanation. For example, if an explanation is inferring that a library is stable because it is open-source, instead of challenging the inference, the challenger asks irrelevant questions like \textit{``Is the library open-source?''}. Another similar error, accounting for 15\% of cases, happens when the challenger asks \textbf{out of scope} questions that lead to incorrect predictions, like asking about the number of developers contributing to an open-source project in the above case.

\nd\ul{\textbf{Error Source - Enquirer (5\%).}} 
Though the Enquirer is expected to split a base-response into an explanation as only one single opinion, sometimes a single explanation still contains multiple opinions (e.g., \textit{``a library performance is great but complex to configure''}). 5\% of misclassifications originate from such \textbf{convoluted challenges}. 

\nd\ul{\textbf{Mixed Error Sources (19\%).}} On 9\% of error cases, CID cannot detect incorrect responses based on inconsistency as ChatGPT \textbf{continuously infers wrong logical reasoning} in all challenge questions. Other \textbf{generic reasons} (9\%) include  unclear information and lack of conclusive opinions in the explanation responses.

\begin{tcolorbox}[
       left=0pt, right=0pt, top=0pt, bottom=0pt, colback=white, after=\ignorespacesafterend\par\noindent]
\textbf{Summary of RQ4.} The best-performing ML model to detect incorrect ChatGPT responses in our benchmark was SVM with an F1-score of 0.75. The major source of misclassification occurred due to the inability of the current similarity measures to handle larger and more convoluted texts for comparison.
\end{tcolorbox}

\subsection{Impact of Challenge Prompts (RQ5)}

\subsubsection{Approach} To understand the separate impact of the basic (or mutation) challenger, we evaluated the performance of CID by only keeping the corresponding 12 similarity attributes related to basic (or mutation) challenges. Moreover, to examine the impact of each of the three basic challenge types (\textit{why}, \textit{how}), we dropped each of them at once and noted CID's performance variance. We used the best performing model (SVM) for this evaluation.

\subsubsection{Results}
\begin{table}[]
    \centering
    \caption{Impact of individual challenge prompts}
    \begin{tabular}{lrr}
    \toprule
\textbf{Scenarios} & \textbf{Accuracy} & \textbf{F1-score} \\ \midrule
With all questions & 0.75 & 0.74 \\
Without \textit{How} questions & 0.75 & 0.75 \\
Without \textit{Really} questions & 0.69 & 0.70 \\ 
Without \textit{Why} questions & 0.70 & 0.71 \\ 
\bottomrule         
    \end{tabular}
    \label{tab:challenge_question_impact}
\end{table}
Table \ref{tab:challenge_question_impact} shows the accuracy and F1-score of the prediction models where different question types are dropped.
When all question types are present in the challenger, the average accuracy for all the prediction models is 0.75. The \textit{How} question seems to have almost no or opposite impact on the prediction since the accuracy remains the same (0.75) without the \textit{How} question, F1-score drops slightly (0.74) without them. \textit{Really} and \textit{Why} questions have higher impact on the prediction than the \it{How} questions. 

\begin{table}[]
    \centering
    \caption{Impact of mutation \& basic challenges}
    \label{tab:mutation_impact}
    \begin{tabular}{lrrrr}
         \toprule
         \textbf{Scenarios} & \textbf{Accuracy} & \textbf{F1-score} \\ \midrule
With Basic and Mutation Challenges & 0.75 & 0.74 \\ 
Without Mutation Challenges & 0.63 & 0.65 \\ 
Without Basic Challenges & 0.69 & 0.69 \\ 
\bottomrule
    \end{tabular}
\end{table}
Table \ref{tab:mutation_impact} shows the prediction model performance in the absence of basic and mutation challenges. When all the challenges are present, the average prediction accuracy is 0.75 whereas if the mutation challenges are excluded, the accuracy significantly drops down to 0.63 (16\% drop from 0.75). However, when only basic challenges are excluded, accuracy does not drop such drastically, it drops to 0.69 (8\% drop from 0.75). Hence, it is obvious that the mutation challenges are more effective than basic challenges in improving the accuracy of the prediction model. 

\subsubsection{Limitation of \textit{How} Challenges} Though we expected better impact from the \textit{how} question types, we find that the prediction model performs better without \textit{how} questions. Moreover, in section \ref{sec:results}, we found that the out-of-scope challenges and the misdirected challenges contribute to 15\% and 18\% of errors respectively. On manual analysis of  the challenge questions in our evaluation study, we found that \textit{how} type challenge questions are contributing to such challenger-based errors. For example, ChatGPT provided a wrong response saying that \textit{``Presence of an open-source community ensures stability''} though there is no such direct causation relation between open-source community and project's stability. To challenge this explanation, the challenger component generated two relevant challenging questions with \textit{why} and \textit{really}:
\textit{``Why active dev community contribute to stability''}
\textit{``Really, is active dev community solely responsible for stability''}.
In response to these questions, ChatGPT agreed that \textit{``No direct answer available''} and \textit{``This is hard to determine''} respectively. Both these responses would show that the original statement from ChatGPT was wrong. However, the challenger component generated an out-of-scope challenge using \textit{how} question type: \textit{``How many developers are contributing to the open source project.''} and ChatGPT responded with \textit{``No information provided''}. Because of this out-of-context question and response, the model got confused and failed to detect the error ultimately. 

\subsubsection{Impact of Question Mutation} Let's provide an example of how mutated questions are improving prediction accuracy. After being asked about the effectiveness of SpaCy in text categorization task (base-question), ChatGPT responded in the affirmative with the following explanation: \textit{``The user is able to train a TextCategorizer model with labeled data pairs using example code \textbf{from Spacy's documentation}''.} However, the user was able to train a model without any reference to spaCy's documentation. Hence, this explanation from ChatGPT was unacceptable. During the basic challenge phase, the challenger asked a question \textit{``How comprehensive is the documentation for training a TextCategorizer model, and are there \textbf{any limitations or potential issues users might encounter?}''} However, in response to this challenge, ChatGPT did not address the limitation part. To the contrary, ChatGPT responded to the basic challenge as \textit{``The conversation suggests that the documentation for training a TextCategorizer model in Spacy is helpful and straightforward... No limitations or potential issues are discussed.''} Then when the challenger component applied question mutation, it added a clause regarding the \textit{limitation} while generating an equivalent mutated question: \textit{``How comprehensive is the documentation for training a textcategorizer model, and are there any limitations or potential issues users might encounter \textbf{without considering how might the limitations of online learning affect the accuracy of the model when adding new entities?}''} As the bold part of this question again points to the limitations of online learning, ChatGPT responds to this mutated challenge this way: \textit{``While the documentation for TextCategorizer training is comprehensive, \textbf{users may encounter limitations and accuracy issues when adding new entities}''.} Thus, this time ChatGPT changes its response from the original response and subsequently in the similarity calculator and prediction model, this was detected as an inconsistent answer and labeled as an incorrect explanation.

\begin{tcolorbox}[
       left=0pt, right=0pt, top=0pt, bottom=0pt, colback=white, after=\ignorespacesafterend\par\noindent]
\textbf{Summary of RQ5.} In CID, the mutation challenge prompts based on the metamorphic relationships have a more significant role to detect inconsistencies than the base challenge prompts.
\end{tcolorbox}

\section{Related Work}
We compare our work with (1) the current techniques developed to test the quality of LLMs and (2) the adoption of LLMs in SE tasks. 

\nd\ul{\textbf{Reliability Issues of LLMs.}} Despite the efficacy of LLMs including ChatGPT, they are often susceptible to generating non-factual statements and even hallucinating facts \cite{feldman2023trapping, zhang2023language, bang2023multitask, galitsky2023truth, dai2023plausible, patil2023gorilla}. Researchers have explored various factors contributing to this issue, such as the quality of training data \cite{wang-2019-revisiting, lee-etal-2022-deduplicating}, source-target divergence \cite{dhingra-etal-2019-handling}, inadequate modeling \cite{aralikatte-etal-2021-focus, feng2020modeling}, and randomness during inference \cite{dziri2021neural}. Several studies leveraged the logit output values as a measure of the model’s `uncertainty' and used them to detect hallucination \cite{varshney2023stitch, jiang-etal-2021-know}. \jRevision{Logit output values refer to the raw probability scores for the generated tokens, which are then transformed using the softmax function and logarithmically scaled to calculate log probabilities.}~However, log probabilities are not a good estimate of hallucination or epistemic uncertainty as they merely indicate the likelihood of specific tokens or ways of expressing a claim \cite{lin2022teaching}. Hence, if a claim can be paraphrased in many different ways, each paraphrase may have a lower log probability, irrespective of its factual correctness. Recently, Azaria and Mitchell leveraged the internal state (i.e., activation values) of the language model to determine the truthfulness of a statement using a separate classifier \cite{azaria2023internal}. \jRevision{Mielke et al. trained a correctness predictor for LLM responses that also directly uses model’s internal representations \cite{mielke2022reducing}.}~This approach requires the internal states of the LLM, which may not be available in a black-box situation (like ChatGPT).~{In contrast, our CID tool is applicable to such black-box settings that do not require any internal state of LLM. Rather it employs iterative prompting and challenges the LLM to reveal any inconsistency that can be indicative of possible incorrectness in the LLM response.} \jRevision{Previous black-box approaches mainly work by generating multiple responses for the same question and capturing the variations in these generations \cite{manakul2023selfcheckgpt, lin2023generating, xiong2023can}. For example, Manakul et al. present SelfCheckGPT, a black-box method for detecting hallucinations in generative LLMs by examining inconsistencies across multiple answers for the same question \cite{manakul2023selfcheckgpt}.~Such methods may overlook redundant misinformation dispersed across multiple responses, masking the presence of hallucinations. Moreover, the variance in responses is subject to the randomness parameters of the language models (as they are generated for the same question). As a result, such approaches are not applicable to deterministic setups.~However, instead of comparing multiple generations for the same question, CID exerts follow-up enquiries and challenge questions to the LLM through iterative prompting to capture inconsistency.}

Similar to other LLMs, ChatGPT also suffers from a lack of consistency \cite{jang2023consistency, jang2022becel}. The model frequently alters its decisions when presented with a paraphrased sentence, revealing self-contradictory behavior.~{Existing research considers the correctness and consistency of the LLMs as two separate objectives \cite{elazar2021measuring}, which may not be correct. ~In this paper, we show that by assessing how and when responses deviate from the previous responses to the same question, we could detect incorrect responses.}

Several studies have explored various approaches to mitigate hallucination in LLMs such as fine-tuning the LLM with human feedback and reinforcement learning \cite{bakker2022fine, ouyang2022training}, factuality-enhanced training \cite{lee2022factuality}, external fact-checking tool employment \cite{gou2023critic}, improved prompting \cite{feldman2023trapping}, and external knowledge incorporation into LLM-generated responses \cite{peng2023check}. ~{Existing research treats detection and mitigation of inaccuracies in LLM chatbots as two separate problems.} ~{Though mitigation of hallucination or factual inaccuracy of LLM does not fall within the scope of this study, our CID tool can warn the users of any possible reliability issue of a given ChatGPT response that might warrant further verification.} 

\nd\ul{\textbf{LLMs for Software Engineering (SE).}} Both specialized and general-purpose LMs/LLMs have emerged as powerful tools for various SE applications. Models like PLBART \cite{ahmad2021unified}, CodeBERT \cite{feng2020codebert}, CodeT5 \cite{wang2021codet5}, CoTexT \cite{phan-etal-2021-cotext} are designed with a focus on addressing the unique challenges and requirements of software development tasks. These architectures rely on large-scale pre-training of both programming (i.e., source code) and natural language (i.e., code comment) data, enabling them to perform various downstream SE tasks. Other noteworthy contributions include Liu et al.'s self-attention neural architecture for code completion and Wei et al.'s dual training framework for code summarization and code generation on GitHub datasets. Additionally, Nijkamp et al.'s CodeGen presents a family of LLMs trained in natural language and programming data that achieved competitive performance in zero-shot Python code generation. In contrast, more general-purpose LLMs (such as GPT-based models \cite{budzianowski-vulic-2019-hello, brown2020language}, Codex \cite{chen2021evaluating}, ChatGPT \cite{ChatGPT_openAI}) have showcased their versatility in tackling a wide range of NLP and software engineering tasks. These models are also pre-trained on vast amounts of natural and programming language data available on the internet. Several studies have leveraged such models for various SE tasks such as code generation \cite{finnie2022robots, khan2023combining, liu2023your}, code repair \cite{xia2023keep, prenner2021automatic, pearce2023examining}, code summarization \cite{khan2022automatic, ahmed2022few, macneil2022generating}. In addition, such models (specifically, ChatGPT) combine conversational capabilities with code-related tasks, allowing programmers to interact with the model. This allows developers to use such models as  programming assistants \cite{tian2023chatgpt}.

We are not aware of any study in SE that focused on understanding practitioners' perspectives on the usage of LLMs or their inaccuracies.~{To fill this gap in SE literature, we first conducted a comprehensive survey of 135 developers and then developed our tool CID. We showed the effectiveness of CID to assess inaccuracies in ChatGPT responses during software library selection.}

\section{Conclusions}
\nd\ul{\textbf{Summary.}} In a survey of 135 SE practitioners, we find that they are trying to use Generative AI-based chatbots like ChatGPT to support their diverse SE tasks. However, they often worry about the truthfulness of ChatGPT responses. We developed a tool called CID to detect incorrect ChatGPT responses. \jRevision{CID repeatedly queries ChatGPT with contextually aligned yet textually divergent questions. For a given question, a
response that is different from other responses (across multiple
incarnations of the question) is likely to be an incorrect response.} In a benchmark study involving a dataset of Q\&As for software library selection, we find that CID can detect incorrect responses from ChatGPT with an F1-score of 0.74-0.75. 

\nd\ul{\textbf{Limitations.}} 
The correctness of ChatGPT responses in our training dataset is determined based on the provided SO post content. While we only collected the accepted answers to ensure their validity, the ground truth may vary among human annotators. For the survey, \textit{\textbf{selection bias}} could arise through the recruitment of survey participants. Our survey sample is quite large (135 industrial SE practitioners), which could mitigate any under/over-representation of participants. Another threat is \textit{\textbf{response bias}}, where participants may provide socially desirable answers or omit critical information. This threat was mitigated by ensuring anonymity and confidentiality in the survey responses. In the tool development, \textbf{\textit{internal validity}} threats relate to biases in the dataset used to train and test the CID tool's components. The training and test sets did not share any common data.  

\nd\ul{\textbf{Generality of CID and Future Work.}} \jRevision{The core principle behind CID revolves around probing for inconsistencies in ChatGPT responses. The probing is done by asking the same question in different ways and inconsistencies are detected in the responses to the question (e.g., when one response diverges from the rest). This philosophy is broad enough to be applied to numerous tasks within the realm of SE and beyond. The current version of CID is tested for ``textual" responses from ChatGPT. Intuitively, such responses could offer answers to questions related to diverse domains. However, the focus of our paper was to determine the trustworthiness of ChatGPT usage in SE tasks and the design and evaluation of CID to support their assessment of the trustworthiness of ChatGPT responses. We picked responses as reviews that are related to software library selection. Our choice is influenced by extensive research in the SE domain in the last few years on the usage of opinions and reviews to facilitate the usage and selection of software libraries~\cite{uddin2017opiner,uddin2019understanding}. However, given the broad nature of the underlying principles of CID, it can be tuned and applied to other SE tasks that could benefit from textual responses from ChatGPT. Other tasks that are mentioned by our survey participants include - (a) Code Analysis and Review, and (b) Exploring Alternative Approaches. In our future work, we aim to assess the generalizability of CID by tuning it for these other similar SE tasks (e.g., code reviews) and by collecting feedback from diverse software professionals during their usage of ChatGPT and our tool CID.}

\section*{Data Availability}
The code and data used for this paper are shared in the replication package \cite{website:replication-package}.

\bibliographystyle{ACM-Reference-Format}
\bibliography{references}

\end{document}